\documentclass[11pt,a4paper]{article}

\usepackage{indentfirst}
\usepackage{amsfonts}
\usepackage{amssymb}
\usepackage[reqno,intlimits]{amsmath}
\usepackage[polish,english]{babel}

\newtheorem{theorem}{Theorem}
\newtheorem{lemma}{Lemma}

\newcommand{\ud}{\mathrm{d}}

\title{Global dynamics of cosmological scalar fields -- Part II}
\author{Andrzej J.~Maciejewski,\\
Institute of Astronomy,
University of Zielona G\'ora\\
Podg\'orna 50, 65--246 Zielona G\'ora, Poland,
(e-mail: maciejka@astro.ca.wsp.zgora.pl)\\[2ex]
Maria Przybylska,\\
Institut Fourier, UMR 5582 du CNRS,\\
Universit\'e de Grenoble I,\\
100 rue des Maths,\\
BP 74, 38402 Saint-Martin d'H\`eres Cedex, France\\
and\\
Toru\'n Centre for Astronomy,
Nicholaus Copernicus University \\
Gagarina 11, 87--100 Toru\'n, Poland, (e-mail:
mprzyb@astri.uni.torun.pl)\\[2ex]
Tomasz Stachowiak,\\
Astronomical Observatory, Jagiellonian University\\
Orla 171, 30-244 Krak\'ow, Poland, (e-mail:
toms@oa.uj.edu.pl)\\[2ex]
Marek Szyd{\l}owski,\\
Complex Systems Research Centre, Jagiellonian University\\
Reymonta 4, 30-059 Krak\'ow, Poland\\
and\\
Astronomical Observatory, Jagiellonian University\\
Orla 171, 30-244 Krak\'ow, Poland, (e-mail:
uoszydlo@cyf-kr.edu.pl)}

\begin{document}
\maketitle

\begin{abstract}
This is the second part of integrability analysis of cosmological models with scalar
fields. Here, we study systems with conformal coupling, and show that apart
from four cases, where explicit first integrals are known, the generic system
is not integrable. We also comment on some chaotic properties of the system,
and the issues of integrability restricted to the real domain.
\end{abstract}

\section{Introduction}

Conformally coupled fields were subject to more rigorous integrability
analysis, as opposed to minimally coupled ones, thanks to the natural form of
their Hamiltonian. As will be shown in the next section, the kinetic part is of
natural form, albeit indefinite, and the potential is polynomial (in the case
of real fields). 

Ziglin proved that the system given by (\ref{ham_orig1}) is not meromorphically
integrable when $\Lambda=\lambda=0$ and $k=1$ \cite{Ziglin:00::}. His methods
were then used by Yoshida to homogeneous potentials which is the case for the
system when $k=0$ \cite{Yoshida:86::,Yoshida:87::,Yoshida:88::,Yoshida:89::}.
Later, Yoshida's results were sharpened by Morales-Ruiz and Ramis
\cite{Morales:99::}, and used by the present authors in
\cite{Maciejewski:00::b} to obtain countable families (restrictions on
$\lambda$ and $\Lambda$) of possibly integrable cases. Also recently,
more conditions for integrability have been given in \cite{Boucher}, although
only for non-zero spatial curvature $k$ and a generic value of energy, that is,
when the particular solution is a non-degenerate elliptic function (in
particular not for zero energy).

Our work shows, that the conjecture of that paper is in fact correct -- as shown
in Section 4 -- the system is only integrable in two cases (with the above
assumptions). We go further than that
and show that for a generic energy value, a spatially flat ($k=0$) universe is
only integrable in four cases. Also, the particular case of zero energy is
analysed and new, simple conditions on the model parameters are found. Finally,
we check that when $E=k=0$, the problem remains open, as the necessary
conditions for integrability are fulfilled.

When it comes to numerical studies of the problem, there are various results,
most notably chaotic
behaviour \cite{Joras:2003dn}, but also fractal structure and chaotic scattering
\cite{Szydlowski:2006qn}. However, it remains unclear whether the widely
exercised complex rotation of the variables changes the system's integrability.
Even for very simple systems it was shown \cite{Gorni} that there might exist
smooth integrals, which are not real-analytic. This question is especially
vital since our Universe clearly has real initial conditions and dynamics.

In what follows, we derive the Hamiltonian system for the conformally coupled
scalar field, and proceed to analyse its integrability. For the basics of the
theory used, and relevant, more detailed literature see the first part
\cite{PartI}.

\section{Conformally coupled scalar fields}

The procedure of obtaining the Hamiltonian is the same as in the case of
minimally coupled fields, only this time the action is
\begin{equation}
    \mathcal{I} = \frac{c^4}{16\pi G}\int \left[\mathcal{R} - 2\Lambda -
    \frac12\left(\nabla_{\alpha}\bar{\psi}\nabla^{\alpha}\psi +
    \frac{m^2}{\hbar^2}|\psi|^2 + \frac16\mathcal{R}|\psi|^2\right)
    - \frac{\lambda}{4!}|\psi|^4\right]\sqrt{-g}\,\ud^4\boldsymbol{\rm x},
\end{equation}
where an additional coupling to gravity through the Ricci scalar $\mathcal R$,
and a quartic potential term with constat $\lambda$
are present, as opposed to the minimal scenario. The cosmological constant
$\Lambda$ and the mass $m$ remain as previously. We keep the
same notation as before and express the involved quantities in comoving
coordinates and conformal time to get
\begin{equation}
    \mathcal L = 6(a''a + Ka^2) - \frac12a''a|\psi|^2 + \frac12 |\psi'|^2a^2 -
    \frac{m^2}{2\hbar^2}a^4|\psi|^2 -\frac{\lambda}{4!}a^4|\psi|^4 - 2\Lambda
    a^4 - \frac12K a^2|\psi|^2,
\end{equation}
from which we remove a full derivative, and introduce new field variables $\psi
= \sqrt{12}\,\phi\exp(i\theta)/a$ to obtain
\begin{equation}
    \mathcal L = 6\left[ \phi'^2 + \phi^2\theta'^2 - a'^2 + K(a^2-\phi^2)
    -\frac{m^2}{\hbar^2}a^2\phi^2 - \frac{\Lambda}{3}a^4 - \lambda\phi^4\right].
\end{equation}
The associated Hamiltonian is
\begin{equation}
    H = \frac{1}{24}\left(p_{\phi}^2 +\frac{1}{\phi^2}p_{\theta}^2 - p_a^2\right)
    + 6\left[ K(\phi^2-a^2) + \frac{m^2}{\hbar^2}a^2\phi^2 + \lambda\phi^4 +
    \frac{\Lambda}{3} a^4 \right].
\end{equation}
We can see that $\theta$ is a cyclical variable because we took the potential
to depend on the modulus of $\psi$ only, so we write a constant instead of the
respective momentum $p_{\theta}=\omega$.

Finally, we express everything in dimensionless quantities, rescaling the
constants, but also the time and momenta (as they are in fact time
derivatives), which results in rescaling the whole Hamiltonian. We do this as
follows
\begin{equation}
    m^2\rightarrow m^2\hbar^2|K|,\quad \Lambda\rightarrow \frac32\Lambda|K|,
    \quad \lambda\rightarrow\frac12\lambda|K|,
    \quad p_x^2\rightarrow 144 p_x^2 |K|,
    \quad H\rightarrow\frac{1}{12\sqrt{|K|}}H,
\end{equation}
when $K\ne0$, and using another of the dimensional constants otherwise.
Thus, eliminating a multiplicative constant, the Hamiltonian reads
\begin{equation}
    H = \frac12\left(p_{\phi}^2 - p_a^2\right) + \frac12\left[k(\phi^2 - a^2) +
    \frac{\omega^2}{\phi^2} + m^2 a^2\phi^2\right] +\frac14\left(\Lambda a^4 +
    \lambda\phi^4\right), \label{ham_orig}
\end{equation}
with $k\in\{-1,0,1\}$ ($K=k|K|$); $\omega$, $\lambda$, $\Lambda$, $m^2
\in\mathbb R$, and $H=0$
in any physically possible setup. However, the addition of radiation component,
which scales like $a^{-4}$ in the original action, leads to a constant in the
Hamiltonian. Thus, it justifies studying energy levels other than zero as well.

We note that for $m=0$ the system decouples, and is trivially integrable as
shown in Appendix A. That is why we will assume $m\ne0$ henceforth. We will
also take $\omega=0$, that is, consider a scalar field equivalent to a real
field after a unitary rotation in the complex $\psi$ plane.

We change the field variables into the standard $q$ and $p$ ones for further
computation, taking
\begin{equation}
\begin{aligned}
    a &= q_1,&\quad p_a&=p_1,\\
    \phi &= q_2,&\quad p_{\phi}&=p_2.
\end{aligned}
\end{equation}
The Hamiltonian is then
\begin{equation}
\begin{aligned}
    H &= \frac12\left(-p_1^2 + p_2^2\right) + V,\\
    V &= \frac12\left[k(-q_1^2 + q_2^2) 
    + m^2 q_1^2q_2^2\right] +\frac14\left(\Lambda q_1^4 +\lambda q_2^4\right).
\end{aligned}\label{ham_orig1}
\end{equation}
\section{Known integrable families}

There are four known cases when the system has an additional first integral,
independent of the Hamiltonian. They were found by applying the so called ARS
algorithm basing on the Painlev\'e analysis \cite{Ablowitz:80::a}. The
following table summarises those results.
\begin{center}
\begin{tabular}{|c|c|c|c|}
\hline
solvability case &$k$&$\Lambda$&$m^2$\\\hline
\rm{(1)}&$0,\pm 1$&$\Lambda=\lambda$&$m^2=-3\Lambda$\\\hline
\rm{(2)}&$0,\pm 1$&$\Lambda=\lambda$&$m^2=-\Lambda$\\\hline
\rm{(3)}&$0$&$\Lambda=16 \lambda$&$m^2=-6\lambda$\\\hline
\rm{(4)}&$0$&$\Lambda=8 \lambda$&$m^2=-3\lambda$\\\hline
\end{tabular}
\end{center}
And the respective integrals of the systems are
\begin{equation}
\begin{split}
(1)&\begin{cases}
H=\dfrac{1}{2}(p_2^2-p_1^2)+\dfrac{k}{2}(q_2^2-q_1^2)-
\dfrac{m^2}{12}(q_1^4-6q_1^2q_2^2+q_2^4),\\
I=p_1p_2+\dfrac{1}{3}(m^2(q_2^2-q_1^2)-3k),
\end{cases}\\
(2)&\begin{cases}
H=\dfrac{1}{2}(p_2^2-p_1^2)+\dfrac{k}{2}(q_2^2-q_1^2)-
\dfrac{m^2}{4}(q_2^2-q_1^2)^2,\\
I=q_1p_2+q_2p_1,
\end{cases}\\
(3)&\begin{cases}
H=\dfrac{1}{2}(p_2^2-p_1^2)-
\dfrac{m^2}{24}(16q_1^4-12q_1^2q_2^2+q_2^4),\\
I=(q_1p_2+q_2p_1)p_2+\dfrac{m^2}{6}q_1q_2^2(q_2^2-2q_1^2),
\end{cases}\\
(4)&\begin{cases}
H=\dfrac{1}{2}(p_2^2-p_1^2)-
\dfrac{m^2}{12}(8q_1^4-6q_1^2q_2^2+q_2^4),\\
I=p_2^4+\dfrac{m^2q_2^2}{3}\left[4q_1q_2p_1p_2+q_2^2p_1^2-(q_2^2-6q_1^2)p_2^2+
\dfrac{1}{12}q_2^2(q_2^2-2q_1^2)^2\right].
\end{cases}
\end{split}
\end{equation}

In this work, we will show, that the above are the only integrable cases, when
$m\ne0$. An important point to note is that there is a complete symmetry with
respect to interchanging $\Lambda$ and $\lambda$. It is a consequence of the
fact, that there exists a canonical transformation of the form
\begin{equation}
\begin{aligned}
    p_1 \rightarrow i\,p_1,& \quad q_1 \rightarrow -i\,q_1,\\
    p_2\rightarrow p_2,& \quad q_2 \rightarrow q_2,
\end{aligned}
\end{equation}
that changes the Hamiltonian to
\begin{equation}
    H = \frac12\left(p_1^2 + p_2^2\right) + \frac12\left[k(q_1^2 + q_2^2) 
    - m^2 q_1^2q_2^2\right] +\frac14\left(\Lambda q_1^4 +\lambda q_2^4\right),
\end{equation}
which is the same after swapping the indices. We shall use this form of $H$,
where the kinetic part is in the natural form, to make the use of some already
existing theorems more straightforward.

\section{Integrability of the reduced problem}

It is possible to give stringent conditions for integrability of the system, by
considering a reduced Hamiltonian. Namely, we can separate potential V into
homogeneous parts of degree 2 and 4:
\begin{equation}
\begin{aligned}
    V&=V_{h2}+V_{h4},\\
    V_{h2}&=\frac12 k\left(q_1^2+q_2^2\right),\\
    V_{h4}&=\frac14\left( -2m^2 q_1^2q_2^2 + \Lambda q_1^2 +
    \lambda q_2^4\right).
\end{aligned} \label{V_parts}
\end{equation}
The following fact is crucial in our considerations: if a potential $V$ is
integrable then its higheest order as well as the lowest order parts are also
integrable. For the proof, see \cite{Hietarinta}
This means that in
our case if $V$ given by (\ref{V_parts}) is integrable then $V_{h2}$ and
$V_{h4}$ must also be
integrable. $V_{h2}$ is the potential of the two-dimensional harmonic
oscillator, thus, it is trivially integrable. However, the homogeneous part
$V_{h4}$ gives strong integrability restrictions for the whole
potential V. We will call $V_{h4}$ the reduced potential and denote it by $\hat V$.

Thus we effectively set $k=0$, and are now in position to exercise known
theorems concerning homogeneous potentials of two variables. In particular the
complete analysis for degree 4 has been completed in \cite{Maciejewski:2005}.

In order to identify our potential with some of the list given in that paper,
we have to check how many Darboux points there exist, and what are the values
of parameters $\Lambda$, $\lambda$ and $m$ that give potentials equivalent to a
particular family.

We say that a non-zero point $(q_1,q_2)=\boldsymbol d$ is a Darboux point of
the potential $\hat V(q_1,q_2)$ when it satisfies the equation
\begin{equation}
    \hat V'(\boldsymbol d) = \gamma\boldsymbol d,
\end{equation}
where $\gamma\in\mathbb C^{\ast}=\mathbb C \setminus\{0\}$. Such a point
corresponds to a particular solution of the form
\begin{equation}
    \boldsymbol q(\eta) = f(\eta)\boldsymbol d,\quad
    \boldsymbol p(\eta) = \dot f(\eta)\boldsymbol d,
\end{equation}
with $f(\eta)$ satisfying the differential equation (for degree 4 potential)
\begin{equation}
    \ddot f(\eta)=-\gamma f(\eta)^{3}.
\end{equation}

As explained in the first part of this paper, particular solutions allow for
studying the variational equations along them, and yield necessary conditions
for existence of additional first integrals. However, the major simplification
discovered in \cite{Maciejewski:2005} is that additionally there is only a finite number
or parameter sets (or non-equivalent potentials) corresponding to integrable
cases.

Following the cited paper's exposition (and notation) we find that our
potential has:
\begin{enumerate}
\item{Four simple Darboux points,} when
$\Lambda(m^2+\Lambda)(m^2+\lambda)\ne0$, and
$\Lambda\lambda\ne m^4$. The only integrable cases are:
\begin{enumerate}
\item{$\lambda=\Lambda=-\frac13m^2$} ($\hat V$ is equivalent to $V_4$),
\item{$\lambda=-\frac83m^2,\quad \Lambda=-\frac16m^2$} ($\hat V$ is equivalent to
$V_5$),
\item{$\lambda=-\frac83m^2,\quad \Lambda=-\frac13m^2$} ($\hat V$ is equivalent to
$V_6$).
\end{enumerate}
\item{Three simple Darboux points,} when $\Lambda=0$, and
$\lambda(m^2+\lambda)\ne0$. There are no integrable families here as
$\mathcal I_{4,3} =\emptyset$.
\item{Two simple Darboux points,} when either $\Lambda=\frac{m^4}{\lambda}$ and
$\lambda(m^2+\lambda)\ne0$, or $\Lambda=\lambda=0$. Again, no integrable
families are present here because $\mathcal I_{4,2}=\emptyset$.
\item{A triple Darboux point,} when $\Lambda=-m^2$. Additionally there is a
simple Darboux point when $\lambda\ne0$. The potential is equivalent to $V_3$
and is only integrable when $\lambda=-m^2$.
\end{enumerate}

There are two immediate implications that follow. Firstly, that the main system
itself with $k=0$ is only integrable in those four cases, and the respective
first integrals are known, as given in the table. Secondly,
as was shown in \cite{Hietarinta}
those cases are the only ones which could be integrable
when $k\ne0$. This happens because the integrability of the full potential
implies the integrability of the homogeneous parts of the maximal and minimal
degree (the latter is trivially solvable in our case).

As the table shows, when the potential is equivalent to $V_3$
(or, to be precise, its integrable subcase) or $V_4$, the second first integral
is known; but $V_5$ and $V_6$ only have known integrals with zero curvature.
And as was shown in \cite{Boucher}, for $k=1$, the values of $\Lambda$ and
$\lambda$ are those of $V_5$ or $V_6$ forbid integrability. This is easily
extended to the $k=-1$ case, since after the change of variables
\begin{equation}
    q_j \rightarrow e^{i\pi/4}q_j,\quad p_j \rightarrow e^{-i\pi/4}p_j,\quad j=1,2,
\end{equation}
we obtain a system with the sign of $k$ changed, but the ratios $m^2/\Lambda$
and $m^2/\lambda$ the same. Thus, concerning the conjecture of the quoted
paper, our results for $k\ne0$ enable us to state, that it is true, when
rational integrability is considered.

However, the above considerations assume that the energy value is generic, so
that the particular solution is a non-degenerate elliptic function. As stressed
in the first part, this does not preclude the existence of an additional first
integral on the physically crucial zero-energy level.

\section{Integrability on the zero-energy hypersurface}

We choose not to investigate the Darboux points, but the variational equations
directly, as they are considerably simpler in this case.
The Hamiltonian equations of (\ref{ham_orig}) are
\begin{equation}
\begin{aligned}
\dot q_1&=p_1,&\quad \dot p_1&=-kq_1+m^2q_1q_2^2-\Lambda q_1^3,\\
\dot q_2&=p_2,&\quad \dot p_2&=-kq_2+m^2q_1^2q_2-\lambda q_2^3),
\end{aligned}
\end{equation}
and they admit three invariant planes as was shown in
\cite{Maciejewski:00::b}. They are
\begin{equation}
\begin{split}
\Pi_k&=\{(q_1,q_2,p_1,p_2)\in\mathbb{C}^4\,|\,q_k=0\wedge p_k=0\},\qquad
k=1,2,\\
\Pi_3&=\{(q_1,q_2,p_1,p_2)\in\mathbb{C}^4\,|\,q_2=\alpha q_1\wedge p_2=-\alpha
p_1\},\qquad \alpha^2=\dfrac{m^2+\Lambda}{m^2+\lambda}
\end{split}
\end{equation}
Obviously two particular solutions are
\begin{equation}
\begin{aligned}
&\{q_1=p_1=0,\,q_2=q_2(\eta),\,p_2=q_2'(\eta)\},&\quad
0&=\frac{1}{2}\left(p_2^2+kq_2^2+\frac{\lambda}{2}q_2^4\right),\\
&\{q_2=p_2=0,\,q_1=q_1(\eta),\,p_1=q_1'(\eta)\},&\quad
0&=\frac{1}{2}\left(+p_1^2+kq_1^2+\frac{\Lambda}{2}q_1^4\right),
\end{aligned}
\end{equation}
and in order to find the third particular solution we make a canonical change
of variables
\begin{equation}
(q_1,q_2,p_1,p_2)^T=B(Q_1,Q_2,P_1,P_2)^T
\end{equation}
where symplectic matrix $B$ has the block structure
\begin{equation}
B=\begin{pmatrix}
\mathbb{A}&\mathbb{O}\\
\mathbb{O}&\mathbb{A}^T
\end{pmatrix},\qquad
\mathbb{A}=\begin{pmatrix}
-b&-a\\
-a&b
\end{pmatrix},\qquad
\mathbb{O}=\begin{pmatrix}
0&0\\
0&0
\end{pmatrix}
\end{equation}
and $a$ and $b$ are defined by
\begin{equation}
a=\sqrt{\dfrac{m^2+\Lambda}{2m^2+\lambda+\Lambda}},\qquad
b=\sqrt{\dfrac{m^2+\lambda}{2m^2+\lambda+\Lambda}}.
\end{equation}
Let us introduce five quantities
\begin{equation}
\begin{split}
\alpha_1&= 2m^2+\lambda+\Lambda,\quad
\alpha_2=3\lambda\Lambda+2m^2(\lambda+\Lambda)+m^4,\quad
\alpha_3=\sqrt{(\lambda+m^2)(\Lambda+m^2)},\\
\alpha_4&=\lambda^2+\Lambda^2-\lambda\Lambda-m^4,\qquad
\alpha_5=\lambda\Lambda-m^4.
\end{split}
\end{equation}
Then, in the new variables, Hamiltonian (\ref{ham_orig1}) has the form
\begin{equation}
H=\frac{1}{2}\left[P_1^2+P_2^2+k(Q_1^2+Q_2^2)\right]+
\frac{1}{4\alpha_1}
\left[\alpha_5Q_1^4+2\alpha_2
Q_1^2Q_2^2+4(\Lambda-\lambda)\alpha_3Q_1Q_2^3
+\alpha_4Q_2^4\right].
\end{equation}
and the Hamiltonian equations read
\begin{equation}
\begin{aligned}
\dot Q_1&=P_1,&\quad
\dot P_1&=-kQ_1-\frac{1}{\alpha_1}\left[\alpha_5Q_1^3
+\alpha_2Q_1Q_2^2+(\Lambda-\lambda)\alpha_3Q_2^3\right],\\
\dot Q_2&=P_2,&\quad
\dot P_2&=-
kQ_2-\frac{1}{\alpha_1}\left[\alpha_2Q_1^2Q_2+3(\Lambda-\lambda)\alpha_3
Q_1Q_2^2+\alpha_4 Q_2^3\right].
\end{aligned}
\end{equation}
Thus, the third particular solution can be seen to be
\begin{equation}
    \{Q_2=P_2=0,\,Q_1=Q_1(\eta),\,P_1=+Q_1'(\eta)\},\qquad
    0=\frac{1}{2}\left(P_1^2+kQ_1^2+\dfrac{\alpha_5}{2\alpha_1}Q_1^4\right).
\end{equation}
Of course, this is only valid for $\alpha_1\ne0$. We investigate what happens
when $\lambda+\Lambda=-2m^2$ at the end of this section.

Normal variational equations along those three solutions (in the position
variables) are
\begin{equation}
\begin{aligned}
    \xi''(\eta) &= \left[-k+m^2 q(\eta)^2\right]\xi(\eta),\\
    \xi''(\eta) &= \left[-k+m^2 q(\eta)^2\right]\xi(\eta),\\
    \xi''(\eta) &= \left[-k - \frac{\alpha_2}{\alpha_1}q(\eta)^2\right]\xi(\eta),
\end{aligned}
\end{equation}
where $q(\eta)$ is one of $\{q_1(\eta),\,q_2(\eta),\,Q_1(\eta)\}$, depending on the
respective particular solution.

We will consider the $k=0$ case first. Changing the independent variable to
$z=q(\eta)^2$, all the NVE's are reduced to the following
\begin{equation}
    2z^2\xi''(z) + 3z\,\xi'(z) - \lambda_i\,\xi(z) = 0,
\end{equation}
whose solution is
\begin{equation}
    \xi(z) = z^{-(1\pm\sqrt{1+8\lambda_i})/4},
\end{equation}
where we have introduced three important quantities
\begin{equation}
    \lambda_1 = -\frac{m^2}{\Lambda},\qquad
    \lambda_2 = -\frac{m^2}{\lambda},\qquad
    \lambda_3 = \frac{\alpha_2}{\alpha_5}.
\end{equation}

Note, that if any of $\Lambda$, $\lambda$ or $\alpha_5$ is zero, the
corresponding particular solution is constant and cannot be used to restrict
the problem's integrability. Thus, we are left with the $E=k=0$ case as
potentially integrable.

When we assume $k\ne0$, or equivalently $k^2=1$, and introduce the same
independent variable $z$ as before, the NVE's read
\begin{equation}
\begin{aligned}
    2z^2(\Lambda z+2k)\xi''(z) + z(3\Lambda z+4k)\xi'(z) + (m^2z-k)\xi(z)=0,\\
    2z^2(\lambda z+2k)\xi''(z) + z(3\lambda z+4k)\xi'(z) + (m^2z-k)\xi(z)=0,\\
    2z^2\left(\frac{\alpha_5}{\alpha_1}z+2k\right)\xi''(z)
    + z\left(3\frac{\alpha_5}{\alpha_1} z+4k\right)\xi'(z)
    - \left(\frac{\alpha_2}{\alpha_1}z+k\right)\xi(z) = 0.\\
\end{aligned}
\end{equation}
First, let us observe that unlike in the previous case, when any of $\Lambda$,
$\lambda$ or $\alpha_5$ is zero, the system is not integrable. This happens,
because the NVE's then becomes the Bessel equation 
\begin{equation}
    s^2\xi''(s)+s\xi'(s)+(s^2-n^2)\xi(s)=0,
\end{equation}
with $n=1$ and in a new variable $s=m\sqrt{z/k}$ (for the first two) or
$s=m\sqrt{-2z/k}$ (for the third). Such equation is known not to posses
Liouvillian solutions \cite{Kovacic:86::}. Together with the results of the
previous section this leads us to the following
\begin{lemma}
System (\ref{ham_orig1}) considered on the zero or generic
energy hypersurface with $k^2=1$ is not integrable when
$\Lambda$ or $\lambda$ is zero. Additionally for $\lambda+\Lambda\ne-2m^2$, it is not
integrable when $\lambda\Lambda=m^4$.
\end{lemma}

Assuming that none of those constants is zero, we rescale the variable $z$ in
the three equations with
\begin{equation}
    z \rightarrow -\frac{2k}{\Lambda}z,\quad z\rightarrow-\frac{2k}{\lambda}z,
    \quad z\rightarrow \frac{2k\alpha_1}{\alpha_5}z,
\end{equation}
respectively, so that all three are transformed into a Riemann P equation of
the form
\begin{equation}
    \xi''(z) +
    \left(\frac{1-\delta-\delta'}{z}+\frac{1-\gamma-\gamma'}{z-1}\right)\xi'(z)
    +\left[\frac{\delta\delta'}{z^2}+\frac{\gamma\gamma'}{(z-1)^2}+
    \frac{\beta\beta'-\delta\delta'-\gamma\gamma'}{z(z-1)}\right]\xi(z) = 0,
\end{equation}
with the following pairs of exponents $(\delta,\delta')$, $(\gamma,\gamma')$,
$(\beta,\beta')$ at its singular points
\begin{equation}
    \left(\frac12,-\frac12\right),\quad
    \left(\frac12,0\right),\quad
    \left(\frac14+\frac14\sqrt{1+8\lambda_i},
    \frac14-\frac14\sqrt{1+8\lambda_i}\right),\qquad i=1,2,3.
\end{equation}
Using Kimura's results on solvability of the Riemann P equation \cite{Kimura:69::}
we check when the difference of the
exponents give us integrable cases, and find that the parameters must belong to
the following families
\begin{equation}
    \lambda_i = \frac{l_i(l_i+1)}{2},\quad l_i\in\mathbb Z,\quad i=1,2,3.
\label{e0k1con}
\end{equation}
These polynomials in $l_i$ are invariant with respect to the change
$l\rightarrow-l-1$, so it is enough to consider non-negative values only.
Furthermore, $\lambda_1$ and $\lambda_2$ cannot be equal to zero, as $m^2\ne0$,
so $l_1$ and $l_2$ need to be strictly positive.

This result can be refined still. First, let us notice, that $\lambda_i$ are
not independent, for $\alpha_i$ are functions of $\lambda$ and $\Lambda$. We
find the relation between them to be
\begin{equation}
    \frac{1}{\lambda_1-1}+\frac{1}{\lambda_2-1}+
    \frac{2}{\lambda_3-1} = -1, \label{relation0}
\end{equation}
with the exception of $\lambda_i=1$. $\lambda_1$ and $\lambda_2$ cannot both be
equal to 1, as that would mean $\alpha_5=0$. If only one of them, say
$\lambda_1$ is 1, then necessarily $\lambda_3=1$ and the only possibly
integrable cases are those with $\lambda_2$ satisfying (\ref{relation0}) with
$l_2\geq2$. The same holds when $\lambda_1$ and $\lambda_2$ are interchanged.

When $l_1$ and $l_2$ are taken to be grater than 1, $\lambda_1$ and $\lambda_2$
are positive, so the relation (\ref{relation0}) requires that $2/(\lambda_3-1)$
is negative. This only happens for $l_3=0$ and it follows that $l_1=l_2=2$,
which is exactly the first known integrable case.
Since $1/(\lambda_1-1)$ and $1/(\lambda_2-1)$are positive and tend to zero
monotonically as $l_i\geq2$ tends to infinity, there are no other solutions,
and no other integrable sets of parameters.

Finally, we turn to see what happens when $\Lambda+\lambda=-2m^2$. This is
equivalent to
\begin{equation}
    \frac{1}{\lambda_1}+\frac{1}{\lambda_2}=2,
\end{equation}
and the same two conditions of (\ref{e0k1con}) because the first two
variational equations can still be used. It is straightforward to check that
the only integer solution of
\begin{equation}
    \frac{1}{l_1(l_1+1)} + \frac{1}{l_2(l_2+1)} = 1
\end{equation}
is $l_1=l_2=1$, which we recognise as the second case of our table.

\section{Conclusions}
Bringing the results of both parts together we can state the following
properties of the system. For the conformally coupled scalar fields:

\begin{theorem}
System (\ref{ham_orig1}) with a generic energy hypersurface is integrable if,
and only if,
\begin{enumerate}
\item $k=0$, and its parameters belong to the four families listed in the
table. Otherwise there exist no additional, meromorphic integral.
\item $k^2=1$, and its parameters belong to the first two families of the
table. Other than that, there exist no additional, rational first integrals.
\end{enumerate}
\end{theorem}
The second part of the above theorem can be strengthened to meromorphic
integrals, although not for all values of the parameters, as described in
\cite{Boucher}.

\begin{theorem}
If system (\ref{ham_orig1}), when restricted to the zero energy hypersurface,
is integrable, then either
\begin{enumerate}
\item $k=0$, or
\item $k^2=1$ and its parameters belong to the first two families of the table,
or
\item $k^2=1$ and one of $\{\lambda_1$, $\lambda_2\}$ is equal to 1, and the
other satisfies the condition (\ref{e0k1con}) with $l_i\geq2$.
\end{enumerate}
Otherwise, the system is not meromorphically integrable. In particular this
means, that for $k^2=1$ if at least one of $\Lambda$ and $\lambda$ is zero, the
system is non-integrable.
\end{theorem}

For the minimally coupled scalar fields, given in \cite{PartI} by system (14):

\begin{theorem}
For $\Lambda=0$, if the system is integrable then necessarily $E=k=0$.
\end{theorem}

\begin{theorem}
When $\Lambda\ne0$, if the system is integrable on a generic energy level then
either
\begin{enumerate}
\item $k=0$ and $9-4m^2/L=(2n+1)^2$, or
\item $9-4m^2/L=(2n)^2$ (regardless of $k$),
\end{enumerate}
for some $n\in\mathbb Z$.
\end{theorem}

\begin{theorem}
For the zero energy hypersurface, and provided that $\Lambda\ne0$, if the
system is integrable then either
\begin{enumerate}
\item $k=0$, or
\item $9-4m^2/L=(2n+1)^2$, $n\in\mathbb Z$.
\end{enumerate}
\end{theorem}

Of course, depending on the properties of the first integrals, we might get
quite different results, and the requirement of meromorphicity or rationality
is still very restricting. As described in the introduction, this leaves open
the question of existence of real-analytic first integrals. Also we recall that
physically the scale factor $a$ cannot even assume negative values, and some
authors argue that when cosmological (instead of conformal) time is used, the
evolution is not, in essence, chaotic \cite{Castagnino:2001xp}. Thus, we would
like to stress that Liouvillian integrability is a mathematical property of the
system, and often the methods used to study it require the complexification of
variables. This means that when restricted to the narrower, physical domain,
the dynamics might be much simpler. And in particular we might be interested in
a particular trajectory whose behaviour is far from generic. It is no surprise
then, that the dynamics of our system when restricted to $a>0$ might appear
regular. It should still be noted that the notion of chaos, although frequently
associated with integrability, has not yet been successfully conflated with it.
And that regular evolution is not necessarily integrable.

\section*{Acknowledgements}
For the second author this research has been partially supported by the
European Community project GIFT (NEST-Adventure Project no. 5006) and by
Projet de l'Agence National de la Recherche ``Int\'egrabilit\'e r\'eelle et
complexe en m\'ecanique hamiltonienne'' N$^\circ$~JC05$_-$41465. And for the
fourth author, by the Marie Curie Actions Transfer of Knowledge project COCOS
(contract MTKD-CT-2004-517186).

\section*{Appendix A. Massless field}

For $m=0$ we can separately solve for each variable, so that we have
\begin{equation}
\begin{aligned}
    E_1 &= -\frac12\dot a^2 - \frac12ka^2 +\frac14\Lambda a^4,\\
    E_2 &= \frac12\dot\phi^2 + \frac12\frac{\omega^2}{\phi^2}+
    \frac12k\phi^2 + \frac14\lambda\phi^4,
\end{aligned}
\end{equation}
with $E_1+E_2=E$ being the total energy. The first of these is immediately
solved, when we substitute $v_1=a^2$ to get
\begin{equation}
    \dot v_1^2 = 2\Lambda v_1^3 - 4kv_1^2 -8E_1 v_1,
\end{equation}
whose solution is
\begin{equation}
    v_1(\eta) = \frac{2}{\Lambda}\wp(\eta-\eta_1;g_2,g_3) +\frac{2k}{3\Lambda},
\end{equation}
with $\eta_1$ the integration constant and
\begin{equation}
    g_2 = \frac43k^2 + 4\Lambda E_1,\quad g_3 = \frac{8}{27}k^3 +
    \frac43k\Lambda E_1.
\end{equation}
Of course, when $\Lambda=0$ the Weierstrass function $\wp$ reduces to
a trigonometric function.

Similarly, for the other variable, we substitute $v_2=\phi^2$ and obtain
\begin{equation}
    \dot v_2^2 = -2\lambda v^3 - 4kv^2 +8E_2v -4\omega^2,
\end{equation}
whose solution is
\begin{equation}
    v_2(\eta) = -\frac{2}{\lambda}\wp(\eta-\eta_2;g_4,g_5) -
    \frac{2k}{3\lambda},
\end{equation}
where
\begin{equation}
    g_4=\frac43k^2+4\lambda E_2,\quad g_5=\frac{8}{27}k^3 + \frac43k\lambda E_2
    +\lambda^2\omega^2,
\end{equation}
and $\eta_2$ is the integration constant. As before, for $\lambda=0$ the
solution degenerates to trigonometric functions.

\end{document}